# Gaming the Arena: AI Model Evaluation and the Viral Capture of Attention

Sam Hind, University of Manchester

sam.hind@manchester.ac.uk

**Abstract**

Innovation in artificial intelligence (AI) has always been dependent on technological infrastructures, from code repositories to computing hardware. Yet industry – rather than universities – has become increasingly influential in shaping AI innovation. As generative forms of AI powered by large language models (LLMs) have driven the breakout of AI into the wider world, the AI community has sought to develop new methods for independently evaluating the performance of AI models. How best, in other words, to compare the performance of AI models against other AI models – and how best to account for new models launched on nearly a daily basis? Building on recent work in media studies, STS, and computer science on benchmarking and the practices of AI evaluation, I examine the rise of so-called 'arenas' in which AI models are evaluated with reference to gladiatorial-style 'battles'. Through a technography of a leading user-driven AI model evaluation platform, LMArena, I consider five themes central to the emerging 'arena-ization' of AI innovation. Accordingly, I argue that the arena-ization is being powered by a 'viral' desire to capture attention both in, and outside of, the AI community, critical to the scaling and commercialization of AI products. In the discussion, I reflect on the implications of 'arena gaming', a phenomenon through which model developers hope to capture attention.

On May 20$^{th}$ 2025, Alphabet CEO Sundar Pichai hosted the Google I/O Keynote – the company's annual product launch event – presenting 'over a dozen models and research breakthroughs' as well as the release of 'over 20 major AI products and features' all since the last I/O event in 2024 (Google, 2025: n. p.). When Google Deepmind's Sir Demis Hassebis – knighted for his services to artificial intelligence (AI) – stepped up to introduce Google's most advanced version of their generative AI model, Gemini 2.5 Pro, he proudly made two announcements. First, not only that it had achieved an 'ELO score' of 1415 for its web development ('webdev') capabilities, but second, and more impressively, that it was now 'no. 1 across all LMArena leaderboards' (Google, 2025: n. p.).

Innovation in AI has always been dependent on key technological infrastructures, from code repositories to computing hardware. Yet industry – rather than universities – has become increasingly influential in shaping AI innovation in recent times (Ahmed et al., 2023). As AI has moved into real-world and commercial environments, some of these infrastructures have come to shape an AI community supercharged by new technological breakthroughs, grandiose policy announcements (US government, 2025), and colossal investment projects (OpenAI, 2025). Beneath the fervour, as practitioners now typically argue, these infrastructures offer the possibility of truly, fairly, and rigorously evaluating the performance of AI models.

Building on recent arguments around the 'competitive epistemologies' (Orr and Kang, 2024) of AI research and the practices of model evaluation (Engdahl, 2024; Luitse et al., 2025), I contend that AI innovation is increasingly being understood as a battleground in which generative forms of AI, or large language models (LLMs), are publicly scrutinized and compared in 'arenas', a term used by the AI community to refer to the game-like environments where such models are compared head-to-head – such as LMArena, the platform mentioned by Hassebis. I contend that the increasingly gladiatorial framing of such work is driving a viral AI culture (Hind, 2025) dependent on the economic cultivation, and capture, of attention (Goree et al., 2024). In short, what I call in this paper the *viral capture of attention*.

The paper begins by considering the measures, communities, and infrastructures of AI innovation, with a specific focus on the last 15 years of development in AI, situating this historical AI research and development (R&D) in relation to recent work in media studies, platform studies, and science and technology studies (STS) on AI. Across this timeframe AI R&D has increasingly taken place within commercial, rather than university, settings, with industry actors driving the transformation of AI models from research fascination to real-world proposition.

The subsequent section provides a background to the role of model evaluation in AI and machine learning culture. Although more prominent in recent years, benchmarks and leaderboards have long structured AI innovation – functioning as mechanisms for AI model developers to capture the attention of the AI community. Organized competitions have also historically structured work in robotics, AI and autonomous settings – facilitated by industry and private funding.

Extending recent work on evaluative AI practices, I then argue that AI model evaluation is undergoing an *arena-ization*, as demonstrated by the rapid rise of user-driven AI model evaluation platforms where AI models are pitted against each other in anonymous 'battles'. Through a technographic study of LMArena, a leading user-driven AI model evaluation platform, I argue that the arena-ization of AI innovation is powered by a 'viral' desire to capture attention driven by evocative framings of gladiatorial battles between AI models. In this, I contend that four themes have been integral to the rapid rise of LMArena: a critique of benchmarks, an identification of the limits of expertise, consideration of how to scale scoring, and the seeking of user attention through participation.

In the discussion, I reflect on the implications of *arena gaming*, an inevitable effect of the arena-ization of AI innovation, where model developers and commercial actors use various techniques that invalidate the principles of fair competition: that model evaluations are independent and that model developers (i.e. players) and model evaluators (i.e. referees or adjudicators) are separate. In other words, I argue that commercial actors are bending the rules of the scientific game model evaluation rests upon.

I conclude that the viral capture of attention in AI is threatening the translation and utility of AI research in the real-world. As I speculate, the consequence of arena-ization might be felt in three ways, through: the modulation of incrementalist approaches to commercial AI innovation, the further production of reduced representations of reality, and the deepening cultural importance of the 'expression of preferences' within our increasingly AI-dependent world.

**AI innovation**

*Measures of AI innovation*

AI R&D has grown significantly over the last few years, according to different measures of innovation. Patents related to AI grew from 3,833 to 122,511 from 2010 to 2023 (Maslej et al., 2025), whilst academic publications in computer science increased from approximately 102,000 to over 242,000 across the same timeframe (Maslej et al., 2025). AI-related uploads to code repositories, a connected measure of development (Cinus et al., 2025), have grown exponentially over the last few years, with public generative AI projects on GitHub jumping from 10,000 to nearly 140,000 since 2020 (GitHub, 2024). In economic terms, corporate investment in AI hit $252.3 billion in 2024, 13 times the total investment in 2014 (Maslej et al., 2025). Private investment in generative AI specifically grew to $33.9 million in 2024, over 8.5 times the figure in 2022 (Maslej et al., 2025).

According to some calculations, there are now over 60 notable generative AI models in existence, with 40 developed in the USA, 15 in China, three in France and further notable models developed in Canada, Israel, Saudia Arabia and South Korea (Maslej et al., 2025). 55 of these notable models are defined as being developed by firms, including Google/Alphabet, OpenAI, Alibaba, Apple, Meta, Nvidia, Anthropic, Mistral AI, ByteDance, and Deepseek (Maslej et al., 2025). Of the top academic institutions involved in developing generative AI models since 2014, Carnegie Mellon University (CMU), Stanford, UC Berkeley, MIT and the University of Washington in the US are joined by Tsinghua University in China and the University of Oxford in the UK (Maslej et al., 2025). Yet as Ahmed et al. (2023) contend, industry's share of the largest AI models had risen from 11% in 2010 to 96% in 2021, with 100% of top-performing models now involving industry partners either alone or in partnership with universities from 2020 onwards. AI innovation, thus, is now dominated by commercial actors.

*Communities of AI innovation*

Within sub-communities of AI, such as the field of machine vision, certain topics dominate, constituting the cutting-edge of AI innovation. Reflecting on submissions to the 2025 iteration of CVPR,[1] the field's leading machine vision conference, organizers identified three 'hot topics' across the 13,008 papers submitted to the conference (CVPR, 2025: n. p.): 3D approaches, image and video synthesis, and multimodal learning (CPVR, 2025). Each of these topics are being driven by a mix of technical and commercial factors, including advances in the computational capabilities of graphics processing units (GPUs) and the rise of commercial chatbots.

Cinus et al. (2025) similarly identify natural language processing (NLP), graph-based learning, and 'data-centric AI' as growing in popularity since 2012, with deep learning topics comprising 30% of academic publications referenced in patents and code repositories from this date until present (Cinus et al., 2025). Whilst Woolgar (1998) understood that scientific and technological innovation was a social process, 'the successful exploitation (or implementation) of new ideas' (Woolgar, 1998: 442), it is evident that innovation in AI and its sub-fields today is subject to the power of commercial actors, such as chip manufacturers (e.g. Nvidia) and AI firms (e.g. OpenAI), who shape the innovation landscape around them.

Most importantly, as this paper contends, AI innovation across these fields is being enabled through specific infrastructures that are beginning to exert considerable influence on AI innovation, the culture of machine learning, and the nature and form of AI products and services themselves.

*Infrastructures of AI innovation*

AI innovation is dependent on a wide array of technological infrastructures, typically open-source and cloud-based. Code repositories such as GitHub or Hugging Face have become a critical infrastructure of AI, in which 'the notion of democratic AI' through open-source projects and initiatives have promoted 'the shift towards ML [machine learning] driven technological innovation' (Burkhardt, 2019: 210), enabling users to 'create, discover and collaborate on ML better' (Hugging Face, 2025: n. p.). As a facilitator of innovation, code repositories enable 'configurative enumeration' (Mackenzie, 2017: 44) such as the process of materializing and 'assetizing' code-related actions (e.g. 'forking').

Open-source machine learning libraries like Meta's PyTorch, and Google/Alphabet's TensorFlow have also enabled Big Tech firms involved in the development of AI products to capture the innovative energies of the developer community (Gray Widder et al. 2024). Similarly, online courses like Google/Alphabet's Machine Learning Crash Course or IBM's Machine Learning with Python enable users to learn the basics of machine learning for free, structuring the development of AI models further (Luchs et al. 2023).

Computational hardware, such as Google's tensor processing units (TPUs), needed for model training is typically styled as important for AI innovation (Ahmed et al. 2023), with 'foundation' or platform models, such as Meta's Open Pretrained Transformer Model, becoming the base upon which further AI products built (Burkhardt and Rieder, 2024). AI platforms, like Google's Vertex AI, similarly promise a 'one-stop shop' (Google Cloud, 2025: n. p.) for enterprise customers, offering businesses the possibility to build and deploy generative AI products quickly and seamlessly.

More broadly, cloud storage infrastructures are critical to AI innovation, with Big Tech firms able to manage and control the emerging AI industry through their cloud divisions, such as Google Cloud Platform (GCP), Microsoft Azure, and Amazon Web Services (AWS) (van der Vlist et al., 2024). In these instances, Big Tech firms play a critical 'intermediary role' (van der Vlist et al., 2024: 4) shaping not only the AI industry itself but other sectors, increasingly reliant on access to cloud storage facilities to power AI products and features.

**Model evaluation in AI and machine learning**

*Benchmarks*

*Model evaluation* has been critical to the development of AI, with benchmark datasets created to help evaluate the performance of AI models. Many popular machine vision benchmark datasets were compiled in the late 2000s and early 2010s, during the foundational stage of machine learning application, such as PASCAL (Everingham, 2010), ImageNET (Deng et al., 2009), KITTI (Geiger et al., 2012) and COCO (Lin et al., 2015). Many comparable natural language processing (NLP) benchmark datasets emerged nearly a decade later, such as GLUE (Wang et al., 2019), MMLU (Hendrycks et al., 2021), GPQA (Rein et al., 2023) and SWE-bench (Jimenez et al., 2024), providing the foundation for the evaluation of contemporary generative AI products.

Many benchmark datasets have been built with industry funding. COCO was built by researchers at Microsoft Research, Google Brain (now DeepMind), and Facebook (FAIR), GLUE involved researchers from DeepMind, GPQA was devised by researchers at Cohere and Anthropic, and SWE-bench acknowledge support from AWS (Amazon), OpenAI, and Anthropic.[2]

Benchmarking, the practice of using benchmark datasets to compare the performance of AI models, has variously been conceptualized as an 'competitive epistemology' (Orr and Kang, 2024), a scientific practice (Thiyagalingam et al., 2022), a process of standardization (Engdahl, 2024) a culture (Campolo, 2025) and a necessary precondition for organization of AI competitions (Hind et al., 2024). Yet as Narayanan and Kapoor (2025: n. p.) understand, 'benchmarks do not measure real-world utility', instead testing models against a set of standardized, statistically-valid proxies designed to stand-in for a real-world context.

As Singh et al. (2025) contend, 'a meaningful benchmark demonstrates the relative merits of new research ideas over existing ones', important for influencing 'research decisions, funding decisions, and, ultimately, the shape of progress in [the] field' (Singh et al., 2025: 2). In this competitive setting, the benchmark dataset becomes an intended level playing field, 'the standard that both brings together particular subcommunities of ML researchers, as well as further enables their "progression" through the competitive and iterative comparison of ML models' (Orr and Kang, 2024: 1876). As the case of Gemini 2.5 Pro demonstrates, new AI models and products gain considerable value from performing – and being shown to perform – well on established benchmarks. Yet as Campolo (2025) writes, 'benchmarking demarcates perhaps less what humans and models are capable of in some abstract sense and more what they actually value...powerfully reduc[ing] these valuations into a single numerical metric on a prediction task' (Campolo, 2025: 35).

*Leaderboards*

In the AI community, leaderboards can be defined as 'any ranking of models or systems using performance-based evaluation on a shared benchmark' (Ethayarajh and Jurafsky, 2021: 2). They have become key battlegrounds for competition between model developers, seen as definitive representations of model prowess and performance. Benchmarks and leaderboards are inseparable: without the leaderboard it is difficult for the AI community to rank model performance. Without the benchmark it is impossible to compare model performance in a standardized manner.

Leaderboards rely on a linear ranking of submissions. As Ethayarajh and Jurafsky (2021) contend, the utility of leaderboards is 'non-smooth': 'leaderboards only gain utility from an increase in accuracy when it improves the model's rank' (Ethayarajh and Jurafsky, 2021: 3). In other words, an increase in performance recorded by one model is only recognized if it beats a higher-ranked model. Any increase in performance that does not supplant a higher-ranked model remains hidden in, or behind, the leaderboard itself. As Espeland (2020) writes, 'rankings are ordinal or interval measures or classifications' (Espeland, 2020: 106) which, through their abstraction enable 'a great advantage in institutionalizing forms of evaluation' (Espeland, 2020: 106). Yet AI leaderboards are not what Espeland calls commensurate rankings, because they do not utilize 'standardized intervals' (Espeland, 2020: 106) - the gap between #1 and #2 in an AI leaderboard will not necessarily be the same between #2 and #3.

Unlike benchmarks, leaderboards do not require laborious construction but may well require diligent management and oversight. In their contemporary form they may be 'automatic' to some degree, with rankings updated after every new submission. Leaderboards might reflect results from organized competitions, remaining open to new submissions long after a competition has ended (cf. Hind et al., 2024), but increasingly represent open competition rather than being collated by single party (i.e. a challenge organizer). Popular AI leaderboards are now populated by entries submitted by model developers – members of the AI community at large who wish to see how their AI model performs against others.

*Competitions*

Organized competitions have typically driven the development of benchmark datasets, with leaderboards critical for comparing performance across competition tasks in fields such as medical imaging and autonomous driving (Hind et al., 2024; Luitse et al., 2024). Competitions such as the PASCAL Visual Object Challenge (2005-2012) and ImageNet Large Scale Visual Recognition Challenge (2010-17) drove early standardization efforts in the machine vision community, and the DARPA Grand Challenges (2004-07) kickstarted the commercial development of autonomous vehicles. Understood as examples of 'innovation contests' (Adamczyk et al., 2012), they have long been seen as vehicles for stimulating, steering, and accelerating, AI innovation. Like other infrastructures of AI innovation, competitions have increasingly been organized, or funded, by Big Tech firms or private benefactors, such as the XPrize (1994-), Hutter Prize (2006-) or Waymo Open Dataset Challenges (2020-).

Whilst benchmarking, leaderboards and competitions have played central roles in the historical development of AI models, the commercialization of AI has drawn greater scrutiny on the role they play in structuring AI model development work (Ethayarajh and Jurafsky, 2021; Goree et al., 2024; Orr and Kang, 2024). This scrutiny has been driven not only by the wider commercialization of AI models and the development of a wide variety of distinct AI products and features, but also by the transformation of *model evaluation* itself, similarly subject to these wider commercialization pressures.

**A technography of LMArena**

In this paper I adopt a technographic approach to study the arena-ization of AI innovation, drawing on foundational work across media studies and STS by Sigaut (1993), Woolgar (1998), Kien (2008) and Bucher (2018) on the ethnography of technology and innovation, and van der Vlist et al. (2024), Hind (2024) and Steinhoff and Hind (2025) on AI topics specifically. In a study of synthetic datasets, Ravn (2025) defines technography as 'a methodology that aims to carve out the broader social implications of technology by closely investigating its technical aspects… [presented in] a wide range of empirical material' (Ravn, 2025: 5).

For Woolgar, technography is a way to 'tease out the congealed social relations embodied within technology' (Woolgar, 1998: 444) as highlighted by Hind (2024). Although Woolgar (1998) writes that technography, 'by analogy with "ethnography"…is an anthropological form of study' (Woolgar, 1998: 444) which, thus, 'require sustained empirical study in technical settings' (Woolgar, 1998: 444), contemporary AI research is presented in various digital, online platforms and forums. As such, as proponents have detailed, much of this empirical study 'in technical settings' can take place in, on, and through, digital environments where the material traces, residues, and impacts of AI innovation are left and felt.

As Rella (2023: 4) also discusses, making sense of AI innovation requires, for example, conducting 'close readings of computer science papers…understood as invaluable sources of epistemological and ethico-political meaning making' (Rella, 2023: 4). These close readings, as Amoore et al. (2023) add, involve 'engaging with a multiplicity of texts' (Amoore et al., 2023: 2), as this is how 'sense making' and 'meaning making' around AI is enabled. This reading patently requires varied modes of engagement, with different AI infrastructures, platforms, and objects requiring different kinds of methodological engagement, functioning as distinct places through which discursive value circulates and accrues.

Accordingly, I examine the various online material – e.g. project blogposts, platform policy documentation, AI interfaces, research papers, Arxiv repositories, and podcasts – that provide evidence of the arena-ization of AI innovation, as demonstrated by the rapid rise of LMArena, which I introduce below. These resources are tabulated and referenced throughout (table 2).

*LMArena*

LMArena is described as 'a neutral, open platform for testing and evaluating AI models',[3] and 'has emerged as a critical platform for live, community-driven LLM evaluation, attracting millions of participants and collecting over 3 million votes to date' (Singh et al., 2025: 42). While competitors that combine 'open' benchmarking and leaderboard features exist, such as the Hugging Face Open LLM Leaderboard and OpenRouter Leaderboard,[4] these are not driven by user evaluations, with the latter preferring to call itself the 'first LLM marketplace' and an 'AI gateway for developers' (OpenRouter, 2025: n. p.).

Arenas refer to the specific categories of tasks models are trained and compared on. For LMArena, these comprise text, webdev, vision, text-to-image, image edit, search, text-to-video and image-to-video.[5] The text arena, for example, comprises 19 categories users can compare models using, including on maths, creative writing, coding and eight different languages.[6] The webdev category, in contrast, offers comparison of the 'performance of AI models specialized in web development tasks like HTML, CSS, and JavaScript'.[7] Users of LMArena simply use the platform like they would any other LLM or chatbot, inputting their desired prompt into the interface. Once submitted, users are presented with two answers, served by anonymous models. Once the user has voted for their preferred output the identities of the models are revealed. Votes are then tallied to create task-specific leaderboards based on the categories mentioned.

LMArena began life as LMSys (Large Model Systems), a collaboration led by scientists at UC Berkeley involving researchers from Stanford, UC San Diego, CMU and the Mohamed bin Zayed University of Artificial Intelligence (MBZUAI) (UAE). In May 2023, LMSys launched Chatbot Arena – the first time reference to arenas is made,[8] and in late 2024, a dedicated site is launched.[9] In early 2025, Chatbot Arena became LMArena, with the launch of a new interface, and the announcement that LMArena had started a company to support the growth of the platform.[10]

In a technical paper published a year after the launch of the original Chatbot Arena project, the developers described it as 'a benchmarking platform for LLMs that features anonymous, randomized battles in a crowdsourced setting'.[11] In May 2025 they announced their first investment, valuing the company at $600million (Metz and Roof, 2025). In September of the same year, LMArena introduced their first commercial product, AI Evaluations, promising to offer 'enterprises, model labs, and developers comprehensive evaluation services grounded in real-world human feedback, showing how models actually perform in practice'.[12]

| Key milestones | Date | Source |
|---|---|---|
| LMSys (Large Model Systems) project launched | 2023 | https://lmsys.org/about/ |
| Vicuna launch announcement | March 30, 2023 | Resource 8 |
| Chatbot Arena launch announcement | May 3, 2023 | https://lmsys.org/blog/2023-05-03-arena/ |
| Chatbot Arena hits 4,700 votes | May 3, 2023 | https://lmsys.org/blog/2023-05-03-arena/ |
| LLM-as-a-Judge paper released | June 9, 2023 | https://arxiv.org/abs/2306.05685v1 |
| Chatbot Arena reaches 240,000 votes from 90,000 users | January, 2024 | Resource 12 |
| Chatbot Arena hits 800,000 votes | March 1, 2024 | https://lmsys.org/blog/2024-03-01-policy/ |
| ChatBot Arena technical paper | March 7, 2024 | Resource 12 |
| Chatbot Arena hits 500,000 votes | March 29, 2024 | https://huggingface.co/spaces/lmarena-ai/lmarena-leaderboard/commit/bf3bf2009524729e529c7a16ce1f485837979946#d2h-120906 |
| LMSys Kaggle Competition launched on 'Predicting Human Preference' ($25,000 first prize) | May 2, 2024 | https://lmsys.org/blog/2024-05-02-kaggle-competition/ |
| LMSys non-profit corporation status established | September, 2024 | https://lmsys.org/about/ |
| Dedicated Chatbot Arena site | September 20, 2024 | Resource 9 |
| LMArena beta launch | April 17, 2025 | Resource 1 |
| LMArena company announcement | | |
| LMArena investment announcement | May 27, 2025 | https://news.lmarena.ai/new-lmarena/ |
| First commercial product: AI evaluations | September 16, 2025 | Resource 3 |
| Image Arena reaches 17,238,698 votes | October 1, 2025 | https://lmarena.ai/leaderboard/image-edit |
| Text Arena reaches 4,222,042 votes | October 8, 2025 | Resource 5 |

**Table 1.** LMArena key milestones.

In the following section I examine five themes integral to the rise of LMArena: a critique of benchmarks, an identification of the limits of expertise, the scaling of scoring, and the seeking of user attention through participation, and how AI models 'do battle'.

**Step into the Arena**

*Beyond benchmarks*

As the cofounders discussed on a podcast hosted by the Silicon Valley-based venture capital firm Andreessen Horowitz, LMArena was conceived as an answer to a technical question: how can we compare, and evaluate, the growing number of new LLMs launched every day?[13] This was not just a general question being asked in the AI community, but one LMArena had a stake in, having developed their Vicuna LLM in early 2023.[14] It was considered so good, alleged the cofounders, that competitors believed it to be a ChatGPT 'wrapper', i.e. simply an interface behind which lay ChatGPT. Yet as they considered at the time, 'evaluating chatbots is never a simple task' and with 'AI chatbots becoming more advanced' LMArena began to argue that 'current open benchmarks may no longer suffice' for evaluating ever-improving LLMs.[15] They concluded that 'developing a comprehensive, standardized evaluation system for chatbots remains an open question requiring further research'.[16]

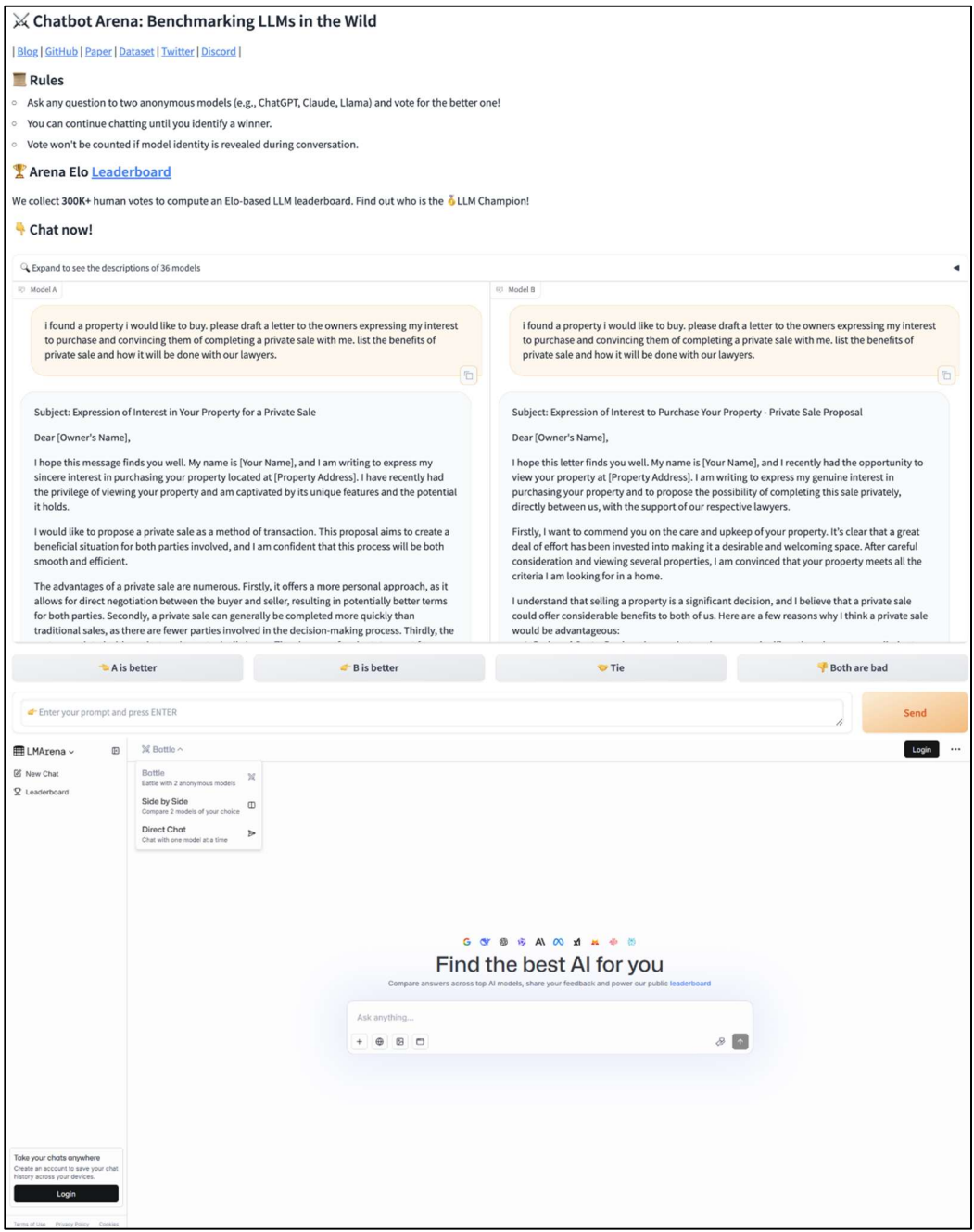

**Figure 1.** Chatbot Arena and LMArena interfaces. Source: Chiang et al. (2024) (top) and author (bottom).

The launch of Chatbot Arena two months later was a direct response to the question of evaluation. At the time, Wei-Lin Chiang, integral to both Vicuna and LMArena, suggested that:

> There was a huge debate internally: when should we release it? How should we evaluate this model? How good is this model really? Does it vibe well, right? Because when we compared it to Llama [Meta's leading open-source LLM], you can feel the difference.[17]

Yet Chatbot Arena wasn't their first attempt to develop a mechanism for evaluating LLMs beyond using benchmarks. As the cofounders were asked, 'what is the right measure of progress [for LLMs]?'[18] As repeatedly suggested, traditional benchmarking arrangements – often structured as straightforward 'exams' such as GPQA – were not considered viable by LMArena.

GPQA, for example, is a dataset 'of 448 multiple-choice questions written by domain experts in biology, physics, and chemistry' (Rein et al., 2023: 1), with experts in these corresponding domains achieving 65% accuracy. Yet, since 2023, the performance of LLMs on GPQA Diamond (a more challenging subset of the whole GPQA dataset) has moved from 31% mean accuracy in June 2023 (GPT-4) to 87% in July 2025 (Grok 4) (Epoch AI, 2025). Their next consideration was whether they could construct an alternative mechanism for evaluating the performance of new LLMs that did not rely on static benchmarks, nor on domain experts compiling challenging exam questions.

*(Automating) expertise*

This question of expertise was of continued interest. Notably, whether expert evaluation of LLMs – often called 'expert labelling' or annotation – would be desirable or even possible. As Anastasios N. Angelopoulos, another cofounder of LMArena, suggested:

> It's great that people do these expert evals [evaluations]…I'm glad we have them. But at the same time, you have to ask yourself: what makes somebody an expert? What are they an expert on? I think the whole world is moving in a direction against experts being the 'be all and end all' of everything. And you know, everybody actually has their own opinions and point of view. There are so many 'natural experts' in the world on all sorts of topics. That they don't necessarily need a PhD in order to be really intelligent…and have valuable opinions.[19]

Whilst this moral questioning of the nature of intelligence and expertise can be considered instructive for LMArena, Ion Stoica, another LMArena cofounder, noted a more practical concern:

> Hard exam followed by hard exam is valuable. No question. The other thing I want to point out is…about having expert labelling. I went to quite a few experts I know, and I respect, and asked them 'would you label?' and almost everyone told me, 'no, I don't have the time'. So there is the question, then, are you really going to get the experts?[20]

After concluding that experts would largely be unwilling to contribute, LMArena decided to consider ways in which they might automate the evaluation process. When OpenAI's ChatGPT-4 launched (in March 2023) they asked 'what if we use it *as a judge*, to do automatic evals [evaluations]?'[21] As reiterated in an evaluation of using LLMs as judges: 'there is a fundamental discrepancy between user perceptions of the usefulness of chatbots and the criteria adopted by conventional benchmarks'.[22] Yet, 'while human evaluation is the gold standard for assessing human preferences [for which LLM/chatbot to use], it is exceptionally slow and costly'.[23]

As they continued, they explained how they considered LLM-as-a-judge to work. Three tests are presented, including what they called 'pairwise comparisons' where 'an LLM judge is presented with a question and two answers, and tasked to determine which one is better or declare a tie'.[24] As they contend, 'LLM-as-a-judge offers two key benefits: *scalability* and *explainability*' reducing 'the need for human involvement, enabling scalable benchmarks and fast iterations'.[25] As the LLMs were asked to provide reasoning for their decisions, the LLM's outputs were deemed 'interpretable'.[26]

*Scaling scoring*

One feature central to LMArena was *scoring*. How might models actually be scored – and scored against other models? The reference for LMArena was the tournament format in which 'you have players play with each other, head-to-head' with 'some number of points to win or lose or tie' after which players are ranked through a leaderboard.[27] Except as Stoica contended, tournaments don't typically involve a carousel of new players:

> And then we thought about other ways in real life how players or teams are ranked – when they don't play each other. Like, when incorporating new players entering into the game. Okay, there are disciplines where this is done like chess, tennis – ATP rankings – and many others. That was the idea, so why don't you do something like an Elo score. So, what do you need? You only need head-to-head. And not everyone needs to play in the same tournament. That's why LMArena has this battle mode in which you have a prompt and…you can pick which one is better.[28]

As suggested,[29] the Elo scoring system, is 'a method for calculating the relative skill level of players', and is the designated system used by the international chess federation (FIDE) for their global chess rankings.[30] An additional 'online' (i.e. real-time) algorithm is also used in the chess rankings to capture presumed player improvement: more recent games count for more than older ones.[31] The Elo system was originally chosen by LMArena precisely because it would enable the ranking of models in a world where new models – i.e. new 'players' in the 'game' – were being released on a daily basis. The Elo system would thus enable pairwise comparisons to be run between any two models, at any time.

Through this process, pairwise win rates between any two models could be predicted. The benefit was that the scoring and ranking of models could easily be scaled. After realizing the need for a more statistically robust approach to model ranking, LMArena replaced the Elo system with the Bradley-Terry model. Angelopoulous joined LMArena on this basis, 'seeing an opportunity to do some interesting statistical modelling and theory'.[32] Like Angelopoulous remarked, such pairwise comparisons are actually *estimations*, where the probability that one model performs better than another is being estimated.

| Resource | Source | Title | Type | Link | Date |
|---|---|---|---|---|---|
| 1. | LMArena | LMArena is growing to support our community platform | Blog | https://news.lmarena.ai/new-beta/ | April 17, 2025 |
| 2. | LMArena | LMArena leaderboard policy | Blog | https://news.lmarena.ai/policy/ | May 27, 2025 |
| 3. | LMArena | New product: AI Evaluations | Blog | https://news.lmarena.ai/ai-evaluations/ | September 16, 2025 |
| 4. | LMArena | LMArena: Prompt. Vote. Advance AI. | Arena | https://lmarena.ai/ | November 11, 2025 (accessed) |
| 5. | LMArena | Text arena | Arena | https://lmarena.ai/leaderboard/text | November 11, 2025 (accessed) |
| 6. | LMArena | WebDev arena | Arena | https://lmarena.ai/leaderboard/webdev | November 11, 2025 (accessed) |
| 7. | LMArena | How it works | Blog | https://lmarena.ai/how-it-works | November 11, 2025 (accessed) |
| 8. | LMSys | Vicuna: An open-source chatbot impressing GPT-4 with 90% ChatGPT quality | Blog | https://lmsys.org/blog/2023-03-30-vicuna/ | March 30, 2023 |
| 9. | LMSys | Announcing a new site for Chatbot Arena | Blog | https://lmsys.org/blog/2024-09-20-arena-new-site/ | September 30, 2024 |
| 10. | Andreesen Horowitz | Beyond leaderboards: LMArena's mission to make AI reliable | Video podcast | https://www.youtube.com/watch?v=YP3Vmh4tYog | May 29, 2025 |
| 11. | LMSys | Chatbot Arena: New models and Elo system update | Blog | https://lmsys.org/blog/2023-12-07-leaderboard/ | December 7, 2023 |
| 12. | Arxiv | Chatbot Arena: An open platform for evaluating LLMs by human preference | Research paper | https://doi.org/10.48550/arXiv.2403.04132 | March 7, 2024 |
| 13. | Arxiv | Judging LLM-as-a-Judge with MT-Bench and Chatbot Arena | Research paper | https://doi.org/10.48550/arXiv.2306.05685 | December 24, 2023 |
| 14. | LMSys | Chatbot Arena: Benchmarking LLMs in the wild with Elo ratings | Blog | https://lmsys.org/blog/2023-05-03-arena/ | May 3, 2023 |

**Table 2.** LMArena resources.

The benefit of the Bradley-Terry model over the Elo system is that it allows for performances to 'converge'. In other words, that over time, it can account for the stabilization of player performance (i.e. that the player eventually stops improving). As each model version doesn't itself change over time – new versions simply become new models, in the eyes of the ranking process – LMArena made the change.[33] As they concluded, 'the core difference between BT [Bradley-Terry] model vs the online Elo system is the assumption that player's performance does not change (i.e., game order does not matter) and the computation takes place in a centralized fashion', rather than games taking place in a decentralized manner (like chess games).[34] As fellow AI start-up Cohere similarly contended, 'unlike dynamic competitors that evolve, LLMs have static capabilities and operate in a time-agnostic context' (Boubdir et al., 2023: 2). Using an Elo scoring system to rank LLMs could lead to inaccurate rankings critical 'given the direct impact of Elo system rankings on both research directions and real-world applications in NLP [natural language processing] as well as its widespread adoption' (Boubdir et al., 2023: 2-3).

The scaling of scoring, thus, was integral to the growth of LMArena as a principal environment in which new models could be easily, and quickly, represented and ranked despite their newness. In other words, that the Bradley-Terry model would best cope with keeping up with the pace of new models being released. Without it, dependent instead on pairwise comparisons between 'static' competitors, new models might fly under the radar. The implication being that new models might not receive the attention they deserved, unable to gain visibility in a crowded field dominated by 'established' models. In such cases the innovative work of those involved would not be duly rewarded – something LMArena could properly enable, they argued, given the switch to a new scoring system.

*Seeking attention, driving participation*

Notwithstanding the limitations to the LLM-as-a-judge approach, including a 'verbosity bias' (where LLM judges are considered to favour longer, inelegant responses) and limited capability in mathematics and reasoning questions, it became clear that crowdsourcing expertise was the preferred evaluation method for LMArena.[35] Firstly, it was considered that this approach could sidestep the expert labour question by not asking domain experts to evaluate LLM outputs at all. Secondly, it could avoid some of the obvious circular complexities of using LLMs to themselves judge LLM outputs, which would no doubt lead to forms of model 'contamination' over longer periods, even if arguably scalable and explainable. But thirdly, calling on ordinary lay users of LLMs could shrink the gap between expert evaluator and actual user, test and real-world to zero. Rather than recruit experts in every conceivable domain – whether science, reasoning, web development or another field altogether – lay users in these spaces could become the adjudicators themselves, arguably best placed to offer their opinion on the best LLM for the task at hand.

But the technical solution to model evaluation arguably also solved another problem for LMArena: how to build a commercial model evaluation *product*. As Stoica suggested:

> What do these companies build? What is their product? Who uses their products? It's not the top experts, it's the laymen. These are the users. That's how OpenAI make their money, and so forth. Then, shouldn't the evaluation take into account the prevalence of these users? The answer is obviously, yes.[36]

This desire eventually culminates in LMArena incorporating as a company,[37] but is entirely dependent on how the platform cultivates a lay userbase to evaluate the models on LMArena itself – arguably otherwise uninterested in participating and sharing their chatbot preferences with LMArena without being lay evaluators too. As Chiang similarly explained:

> When you build a product, an AI product, you want to know how people use it. You want to understand why users prefer this [model] over that. In order to collect that kind of data you have to build a product first, right? And then, as we know, over the past few years people have been building very different kinds of applications on top of AI, beyond just chatbots. These days people are applying these models to…coding, and then more tools that use agentic behaviour etc. How can we capture all these use cases?[38]

This is similarly reiterated by Angelopoulos:

> That's why it's important to us that we continued to grow our community, and we get a diverse community, of all different people, experts, non-experts, artists, scientists, different languages, everybody under the sun. We want them to come to this platform to *express their preferences*. (Emphasis added)[39]

When it comes to the question of attention, then, this can be said to be the driving force behind LMArena itself: an attempt to endlessly proliferate, and constantly diversify, user preferences on the platform. For model developers the same dynamic is true also: 'in order to do well on the Arena, new users need to come and vote for your model. That's it. It means that *users like it*.'[40] It is not simply that this sentiment reflects a desire, and need, for those in the AI industry to promote and drive interest in their work. As the LMArena cofounders articulate, attention is encoded in the (open) arena mechanism itself: in order to become visible on the platform, and in the hope of topping leaderboards on the platform, model developers are pressed into inserting themselves into the logic of model attention, in order to drive up user votes for their models. On the other side, if users want to see their preferences count, they need to participate also. More model users mean more model votes and – if all works out well – higher arena leaderboard positions and comparative scores.

*Model battling*

The cumulative effect of these decisions – a critique of benchmarks, the limits of expertise, the scaling of scoring, and the seeking of user attention through participation – is the construction of the arena of AI innovation. The crucial driver of the ongoing arena-*ization* of model evaluation is what LMArena call *model battling*. Here, statistical terminology around 'pairwise comparisons' was usurped by a more evocative discursive framing where models began to be said to be engaged in gladiatorial 'battles'.

The shift in terminology was embodied in the redesign of the LMArena interface, after which users were informed they could 'battle with two anonymous models' in order to compare their respective capabilities.[41] LMArena users, it was wagered, did not need to know they were contributing to an exercise in statistical prediction. All LMArena users were presumed they wanted to know was that they were participating in the great spectacle of the AI battle, hosted in the grandest arena of them all, LMArena.

Where LMArena differed again was their stated promise to 'advance AI development and understanding through live, open, and community-driven evaluations'.[42] In such instances, they intended to evolve the AI leaderboard beyond a purely open submission ethos where leaderboards might be organized and policed by a certain group (i.e. the research team who created the accompanying benchmark dataset) and instead were organized in an open fashion by the AI user community itself. As they contended:

> Our periodic leaderboard and blog post updates have become a valuable resource for the community, offering insights into model performance that guide the ongoing development of LLMs. Our commitment to open science is further demonstrated through the sharing of user preference data and one million user prompts, supporting research and mode improvement.[43]

Model battling is thus critical to how LMArena scaled, and commercialized, their platform. In the remainder of the paper, I reflect on how this arena-ization of model evaluation is necessarily cultivating the *gaming* of AI innovation, furthering the increasingly viral culture of machine learning more generally.

**Discussion: Gaming the arena**

The effect of the arena-ization of model evaluation is *arena gaming*, increasingly being discussed by the AI community and wider tech press (Whitman, 2025). In the strictest sense, this might be understood as a form of 'overfitting': AI models that perform well on test data but poorly in the real-world, typically as a result of being trained on less diverse datasets. Yet more than simply a case of overfitting, arena gaming involves the optimization of AI models purely for model battling purposes. The aim of optimizing an AI model or product for such a purpose is to climb relevant leaderboards on arena-style platforms – to *dominate the arena*.

More fundamentally, echoing Goree et al. (2024) the objective is to 'leverage [the] highly specialized early attention' (Goree et al., 2024: 8) that can be accrued when developers submit their models to popular leaderboards. Whilst this 'attention' circulates through multiple locations and texts, from GitHub and arXiv pages to conference proceedings, as Goree et al. (2024) note, the leaderboard quantitively distils it, promising an easily-interpretable record of model performance. In this section I discuss how arena gaming is being exhibited through three practices: bespoke comparisons, proprietary privileging, and the compromising of independence.

*Bespoke comparisons*

As Ethayarajh and Jurafsky (2021) have proposed, some of the impacts of these mechanisms can be mitigated through specific design techniques. Regarding arena gaming, they suppose that one option might be the development and proliferation of 'a leaderboard for every type of user' (Ethayarajh and Jurafsky, 2021: 5), supposing that the issue lies in the limited number of performance metrics any one leaderboard employs. Here, users could toggle certain requirements and demands, not purely based on typical brute-force performance metrics, but an array of other functions, such as lightweight models, in order to gain greater real-world value for users who need to make decisions about which AI model to deploy based on e.g. energy efficiency. As they contend, in reference to NLP leaderboards specifically:

> Given the diversity of NLP practitioners, there is no one-size-fits-all solution; rather, leaderboards should demand transparency, requiring the reporting of statistics that may be of practical concern. Equipped with these statistics, each user could then estimate the utility that each model provides to them and then re-rank accordingly, effectively creating a custom leaderboard for everyone. (Ethayarajh and Jurafsky, 2021: 6).

A new version of LMArena, mentioned previously, makes use of generative AI to offer a version of Ethayarajh and Jurafsky's (2021) suggestion. Users interested in using an AI model for a specific purpose compose a prompt outlining their needs. Once issued, LMArena will provide the user with a performance breakdown of two anonymous models. Once assessed, the user chooses a preferred model for the assigned task, and their decision 'helps shape the public AI leaderboards' on the wider LMArena platform.[44]

A similar approach by Reus et al. (2024) proposes designing benchmarks for 'downstream utility' (Reul et al., 2024: 3) in real-world situations and on real-world tasks. As Narayanan and Kapoor (2025: n. p.) suggest, 'by focusing heavily on capability benchmarks to inform…understanding of AI progress, the AI community consistently overestimates the real-world impact of the technology'.

*Preferential privileging*

Another insight from Singh et al. (2025) is that Big Tech developers of AI models including Meta, Google/Alphabet, OpenAI and Amazon are offered 'preferential treatment' (Singh et al., 2025: 4) on community model comparison platforms like LMArena, by 'allowing select providers to test many submissions privately in parallel' (Singh et al., 2025: 4) to public leaderboards without results being shared publicly. In essence, providing some providers with the ability to endlessly test behind the scenes before going public with their benchmark scores. In addition, 'proprietary model providers collect significantly more test prompts and model battle outcomes than others' (Singh et al., 2025: 4) with Google/Alphabet and OpenAI receiving an estimated 19.2% and 20.4% of all test prompts on LMArena. The result of both aspects – private test submissions, significant share of model prompts shared – is that developers of proprietary AI models are significantly privileged in comparison to other model developers, in the emerging model arena environment.

This desire is so great that arena gaming is being used clandestinely by some AI firms to bend the principles of benchmarks and leaderboards themselves, as happened when OpenAI failed to disclose its funding of FrontierMath, a leading maths benchmark developed by Epoch AI (Bastian, 2025). Only a fifth revision of the accompanying research paper 'gratefully acknowledge[d] OpenAI for their support in creating the benchmark' (Glazer et al., 2024: 1) where previous versions had remained silent. OpenAI's new o3 model had achieved a 25.3% success rate on FrontierMath, when other rival models had failed to cross a 2% threshold (Bastian, 2024), wildly deviating from the incremental reality of commercial AI development to date.

More than simply a unnamed funder, it was revealed by Epoch AI that OpenAI had commissioned them to 'produce 300 advanced math problems for AI evaluation that form the core of the FrontierMath benchmark' and in doing so, had 'access to the problems and solutions, with the exception of a holdout set' (Besiroglu and Sevilla, 2025).[45] Whilst it is unknown whether OpenAI model developers had access to these problems and solutions whilst building o3 (thereby improving the possibility that the model would score well on the new benchmark), OpenAI's role of both benchmark commissioner and model developer not only crosses evident ethical red lines, but also constitutes a fairly obvious case of arena gaming.

The developers of LMArena also, for instance, are not unaware of its limitations, acknowledging that 'although our user base is extensive, we anticipate that it will primarily consist of LLM hobbyists and researchers who are eager to experiment with and evaluate the latest LLMs', not only likely to 'result in a biased distribution of users' but thanks to fierce competition in the AI innovation field – the desperate seeking of model fame and glory – to both covert and overt attempts to game the model evaluation process.[46]

*Compromised independence*

The version of model evaluation offered by LMArena is built on a presumption of independence: that model comparisons are carried out by the 'community' (i.e. LMArena users), that the ranking of 'battles' is calculated through a scoring system implemented by LMArena, and that model developers are distinct from either of these two groups. In this game-like arrangement, model developers are 'managers', LMArena is the 'administrator', the models are the 'players', and the model evaluator system is the 'referee' appointed by, but crucially also independent from, the administrators themselves. As Angelopoulous explains:

> I think if [LMArena] came from an industrial lab, people would always have questions about 'oh well, are these people also training a model, and what's their incentive?' and so on. But the reality is that we were just students, we were doing this in order to evaluate models – and that came from a scientific perspective. And I think that's something that people see when they look at us and builds a lot of trust in our business'[47]

Here, presumed scientific 'neutrality' is framed as beneficial to how they can now promote their commercial activity. Yet, as this paper has documented, as LMArena has moved into a more commercial setting this presumed independence is struggling to be maintained. With private testing, preferential privileging, hidden funding arrangements, and the constant drive to seek attention, the 'purity' of the pursuit of scientific knowledge is under threat.

**Conclusion: On the viral culture of model evaluation**

The arena is becoming a distinct battleground for innovation in the AI community, aided by the rise of user-driven model evaluation platforms. The value of established infrastructures of model evaluation is multiplied by their connection to each other: without benchmarks models cannot be compared, and without leaderboards models cannot be ranked. Increased interest in participatory model evaluation, however, has arguably led to arena gaming: where models are built only for the AI arena, rather than the real world, driven by the desire to capture attention – like other computational and 'social' media before it (Simon, 1971; Goldhaber, 1997; Franck, 2018).

This paper has sought to articulate the effect of the arena-ization of AI innovation by offering a sense of how the AI community is not only rapidly scaling and commercializing AI products and features, but also how it is attending to various implications that the rapid scaling and commercialization entails. Yet the AI community is far from united, as one Reddit poster recently suggested: 'I'm starting to think AI benchmarks are useless'.[48] In this conclusion I consider three implications of what I refer to as the viral culture of model evaluation: the modulation of incrementalism, the tokenization of worlds, and the power of preferences.

The dominant approach to AI innovation is *incrementalism*, driven by the 'pragmatic necessities of commercial AI development' (Hind et al., 2024: 1669). As such AI model evaluation, also *temporalizes* AI innovation by 'set[ting]… rankings in motion over time, a movement that produces its own legitimation through incremental improvements' (Campolo, 2025). By setting 'rankings in motion' benchmarks, leaderboards and arenas engender a sensibility where the primary goal may be understood by practitioners not as building a workable or viable 'real world' model, but as keeping up with – with the hope of surpassing – leading models.

Yet as Hind (2025) has considered, AI is a sport of 'conflicting codes' in which the 'game of building the "best" model is fraught with unknowns' (Hind, 2025: 2). Here the intrinsic moral or social value of building a particular AI model – e.g. the recognition of cancerous legions in x-ray images or the summarization of a video meeting - does not itself compel researchers to work on solving such problems. In 'the increasingly viral world of machine learning', as Hind (2025: 4) further contends, 'interest is piqued through the attractive properties of any one model, benchmark, task or competition' as opposed to the 'inherent properties of a domain' (Hind, 2025: 4) flattened into datapoints – and then tokens – for AI processing (Salvator, 2025). The viral nature of contemporary generative AI, thus, inevitably accelerates *tokenization* in all senses of the term: as a *reductive representation* of reality, as a *unit of value*, and as a *symbolic gesture*.

As AI models, products and features continue to be developed, the scrutiny of how AI innovation is conducted, compared, ranked and valued is likely to only increase. The infrastructures of AI innovation work, centred around the phenomenon of arenas, are undoubtedly likely to change, morph and develop as further innovations in AI products continue. As Franck (2018) considered, 'circulation figures and audience ratings are measures of the attention drawn by a medium as a medium' (Franck, 2018: 9). The circulation figures valued by newspapers, and the audience ratings chased by television networks, are not the metrics – quite clearly – valued by developers of new AI models. But neither have researchers been able to rely on the technical metrics developed within research settings, largely unable to capture the real-world panoply of tasks, contexts, and settings in which generative AI tools are being employed.

LMArena has become a critical case in how AI performance evaluation is itself scaled far beyond the AI research community and into the wider – presumed – *user* community itself. What I have argued here, then, is that emerging platforms – or arenas – like LMArena enable the 'measures of the attention drawn by a medium as a medium' in ways that move beyond the ways attention has been used to design AI models themselves (Vaswani et al., 2017; Mackenzie, 2025). Chasing and scaffolding attention, thus, is becoming a principal aim within the AI industry, and a key to an industry obsession in 'scaling' AI technologies themselves (Pfotenhauer et al., 2021).

Through the model arena the value and utility of any given model, regardless of task, context, and setting, can be determined. This determination, as the founders of LMArena have articulated in various fora, is dependent upon the ever-growing aggregation of the *expression of preferences* accumulated in, and through, the arenas in which models do battle.

Understanding how these mechanisms are used, abused, disavowed, and critiqued by the AI community to enable the development of AI models will be crucial, as well as the undoubted consequences of their utilization for science, society, and culture.

**Acknowledgements**

I would like to thank organizers and attendees of the PLAMADISO workshop at the Weizenbaum Institute in Berlin in June 2025 for feedback on an early version of this article. I would also like to similarly thank attendees of the AHCP seminar series in October 2025 for further feedback on a work-in-progress version of the article.

**Declaration of conflicting interest**

The author declared no potential conflicts of interest with respect to the research, authorship, and/or publication of this article.

**Funding statement**

No funding.

**Notes**

1. CVPR stands for Computer Vision and Pattern Recognition.
2. https://www.swebench.com/
3. Resource 1.
4. https://huggingface.co/spaces/open-llm-leaderboard/open_llm_leaderboard; https://openrouter.ai/rankings
5. Resource 4.
6. Resource 5.
7. Resource 6.
8. Resource 11.
9. Resource 9.
10. Resource 1.
11. Resource 12.
12. Resource 3.
13. Resource 10.
14. Resource 8.
15. Resource 8.
16. Resource 8.
17. Resource 10.
18. Resource 10.
19. Resource 10.
20. Resource 10. Stoica is a professor of computer science at UC Berkeley and a cofounder of multiple tech projects including the open-source analytics engine Apache Spark.
21. Resource 10.
22. Resource 13.
23. Resource 13.
24. Resource 13.
25. Resource 13.
26. Resource 13.
27. Resource 10.
28. Resource 10.
29. Resource 14.

30. https://ratings.fide.com/
31. Resource 11.
32. Resource 10.
33. 'However, it is worth noting that the hosted proprietary models may not be static and their behavior can change without notice'. Resource 11.
34. Resource 12.
35. Resource 13.
36. Resource 10.
37. Resource 1.
38. Resource 10.
39. Resource 10.
40. Resource 10.
41. Resource 4.
42. Resource 2.
43. Resource 2.
44. Resource 7.
45. A holdout set is a portion of data left out of a principal training dataset for testing and validation purposes.
46. Resource 12.
47. Resource 10.
48. https://www.reddit.com/r/LocalLLaMA/comments/1i4vwm7/im_starting_to_think_ai_benchmarks_are_useless/